\documentclass[conference]{IEEEtran}

\usepackage{amsmath}
\usepackage{graphicx}
\usepackage{listings}
\usepackage{xcolor}
\usepackage{amsthm}
\usepackage{url}

\newtheorem{example}{Example}

\lstdefinestyle{promptstyle}{
  basicstyle=\scriptsize\ttfamily,
  breaklines=true,
  postbreak=\mbox{\textcolor{red}{$\hookrightarrow$}\space},
  frame=single,
  framerule=0.5pt,
  rulecolor=\color{gray!30},
  backgroundcolor=\color{gray!5},
  captionpos=b
}

\usepackage{pgfplots}
\pgfplotsset{compat=1.18}

\begin{document}
\title{Data-aware candidate selection in NL2SQL translation via small separating instances}
%
%
\author{
\IEEEauthorblockN{Stanislav Kikot}
\IEEEauthorblockA{Huawei Technologies Ltd}
\and
\IEEEauthorblockN{Alexander Shulgin}
\IEEEauthorblockA{HSE University}
\and
\IEEEauthorblockN{Yanwei Xu}
\IEEEauthorblockA{Huawei Technologies Ltd}
}
%
%

%
%
\maketitle              
\begin{abstract}
We propose a data-aware candidate selection method for NL2SQL translation 
based on separating instances and provenance. 
We implement this approach and evaluate it against three natural baselines 
on a subset of BIRD-DEV. 
Experiments show that our method consistently outperforms 
baselines when only two or three candidates are given
and no consistency score is available. The code of our prototype 
can be found at \url{https://github.com/staskikotx/SISelection}.
\end{abstract}
\section{Introduction}

Natural Language to SQL (NL2SQL) translation has emerged as a critical frontier in the broader effort to democratize access to structured data. As enterprises accumulate vast repositories of information in relational databases, the ability to query these systems remains largely confined to users with technical expertise in SQL. This creates a significant bottleneck in data-driven decision-making, particularly for non-technical stakeholders such as business analysts, product managers, and customer support personnel.  Natural language interfaces aim to bridge this gap by enabling users to express information needs in everyday language, which are then automatically translated into executable SQL queries. 

\paragraph{BIRD Challenge} To better reflect the complexity of real-world enterprise databases, the BIRD (BIg Bench for Large-Scale Database Grounded Text-to-SQLs) was introduced as a large-scale, cross-database NL2SQL evaluation framework featuring realistic sche\-mas, rich documentation, and complex queries that involve multi-table joins, aggregations, and implicit domain knowledge. The primary evaluation metric is execution accuracy, ensuring that the generated SQL when run on the target database produces the same result as the `gold' answer. In spite of the criticism \cite{jintext}, 
we support this approach, because it encourages NL2SQL systems to take
into account \emph{implicit conventions and constraints} from real-world 
databases.

\paragraph{Candidate Selection} In modern LLM-based NL2SQL systems, accuracy is often limited not by the ability to generate correct SQL, but by the inability to select it. To improve robustness, most systems generate multiple candidates using stochastic decoding or multiple prompts, leveraging the inherent randomness of LLMs to explore diverse query hypotheses. While this increases the likelihood that a correct SQL query is produced, the final choice typically relies either on self-consistency or on specifically trained LLM-based classifiers and often fail to identify the best candidate.
For example, on the BIRD benchmark, Alpha-SQL with Qwen-7B-Instruct produces the correct query within the top 24 candidates in 84\% of cases, but achieves only 65\% execution accuracy
\cite{li2025alpha}. Enhancing selection accuracy represents an opportunity to boost overall system performance. Other possible applications of the candidate selection problem include 
solving multiple-choice NL2SQL tests (to our knowledge they do not exist, but could)
and automated ambiguity resolution based on pragmatics and the database content
\cite{zhang2025clear}.

\paragraph{Separating instances} The abstract problem
\begin{quote}
given queries $Q_1$ and $Q_2$, build a database in\-stance $D$ on which $Q_1$ and $Q_2$ give different answers,
\end{quote}
is the flip side of the well-known theoretical problem of deciding if two SQL queries are equivalent. Its state-of-the-art solution VeriEQL \cite{he2024verieql} takes into account schema constraints and uses symbolical query evaluation and Z3 solver \cite{de2008z3} to avoid the search assuming a fixed number of rows in each table. This tool was used in \cite{yang2025automated} for semiautomatic detection and correction of human-made errors in NL2SQL benchmarks. 
However,  constructing separating instances `from scratch' does not increase the quality of
candidate selection because it ignores \emph{implicit constraints} on the target database
(such as `patient.name cannot be NULL' and also many more complex ones) which play a crucial role in answer verification.

Thus, a more relevant to us version of this problem is
\begin{quote}
given queries $Q_1$ and $Q_2$ and a database $D$ such that $Q_1(D) \neq Q_2(D)$, 
find a small subset $D'$ of $D$, such that $Q_1(D')  \neq Q_2(D')$.
\end{quote}
This problem was studied in context of SQL tuition to the students
\cite{miao2019explaining}. In this setting the instance $D'$ can be regarded as an explanation of why the student’s 
query $Q_1$ is incorrect in comparison to the model answer $Q_2$. Thus, an educational system RATEST \cite{miao2019explaining}, developed at Duke University, reveals $D'$ to a student who submits incorrect $Q_1$ without revealing the model answer $Q_2$ and encourages them to correct $Q_1$ based on this information.

\paragraph{Provenance and ProvSQL} In database research, provenance 
\cite{buneman2007provenance,green2007provenance} denotes the derivation metadata that track the origin and query-level transformation of relational tables.
Two most popular RDBMS extensions with provenance support 
are GProM \cite{arab2018gprom} and ProvSQL \cite{sen2025provsql}. They both are based on query
rewriting and support different subsets of SQL.

\paragraph{Other related research} In \cite{fan2025grounding} a provenance engine is used
for explaining and improving candidate selection in NL2SQL. 
In \cite{klopfenstein2026spotit+}, given
two queries, the system tries to construct a separating instance which is not necessary
extracted from the given database instance but must be similar to it in terms of
range constraints on values. In \cite{li2026dpctrainingfreetexttosqlcandidate} so called
`minimal distinguishing databases' are constructed using LLMs and Python.

\paragraph{Candidate selection methods used by BIRD leaders}

Out of top 30 submissions on BIRD leader board only 17 are accompanied with research papers or open-source code, out of which only 13 are different. Let us go through them.
Authors of \cite{shkapenyuk2025automatic} use simple heuristics and majority voting (also knows as self-consistency) for candidate selection. After removing queries violating certain simple rules all remaining candidates are evaluated on the model database, then they are grouped according to the answers they give, and a representative from the largest group is picked. Authors of \cite{pourreza2024chase} and \cite{liu2025xiyan} use a fine-tuned LLM for binary candidate selection. Authors of \cite{bird_sql} and \cite{csr_sql} do candidate selection by submitting queries, schema, general knowledge and candidate execution results to a general LLM. Authors of
\cite{sheng2025csc} and \cite{pourreza2025reasoning} use reinforcement learning and do not have a distinct candidate selection module. Authors of \cite{donder2025cheaper} combine majority voting with a general LLM voting. Authors of \cite{xie2025opensearch}, 
\cite{yang2026memo}, \cite{li2025omnisql}, \cite{maamari2024death},  and \cite{li2025alpha} rely on self-consistency. 
Authors of \cite{li2025deepeye} use a combination of a consistency score and a multiple 
binary selection by a general-purpose LLM.
Authors of \cite{talaei2024chess} present an original selection strategy based on so called ‘unit tests’ generated by an LLM to highlight the difference between the candidates and pick a candidate that passes the greatest number of these unit tests. Their unit-tests are similar to our separating instances, however, they are constructed by LLMs and do not take $D$ into account.
The authors of \cite{wang2025agentar} train a custom outcome reward model for candidate selection in context of NL2SQL.

\paragraph{Imperfection of existing methods for candidate selection}

The disadvantage of existing candidate selection strategies is that neither of them 
takes into account the model database $D$ fully. Even if the selection is carried out by an LLM which sees the results of execution of $Q_1$ and 
$Q_2$ on $D$, the LLM does not have access to $D$ due to its prohibitively large size. For example, suppose that we want to select between two candidates
\begin{quote}
SELECT s.district FROM satscores AS ss INNER JOIN schools AS s ON ss.cds = s.cdscode WHERE s.statustype = 'Active' ORDER BY ss.avgscrread DESC LIMIT 1\hspace{4.2cm} $(Q_1)$
\end{quote}
and
\begin{quote}
SELECT satscores.dname FROM satscores INNER JOIN schools ON satscores.cds = schools.cdscode WHERE schools.statustype = 'Active' ORDER BY satscores.avgscrread DESC LIMIT 1 \hspace{1cm} $(Q_2)$
\end{quote} for the question “Which active district has the highest average score in reading?” How can the LLM tell which of them is correct without seeing $D$ 
but seeing $Q_1(D) =$ \{($'$Santa Cruz County Office of Education$'$)\} and 
$Q_2(D) = $ \{($'$Palo Alto Unified$'$)\} ?

\paragraph{Contribution} We suggest to provide the selection agent, which deals with candidates $Q_1$ and $Q_2$, with a separating instance $D’$ which is extracted from $D$ by executing $Q_1$ and $Q_2$ on an RDBMS with a provenance support and analyzing the difference between the rows that contributed to $Q_1$ and to $Q_2$.  Given such $D’$, we can select between $Q_1$ and $Q_2$, for example,  by evaluating them on $D’$ using a standard RDBMS, then evaluating the query in natural language on $D’$  using a general LLM that sees the NL question and all database metadata, 
and comparing these answers. This results in a novel method for candidate selection in NL2SQL
translation based on separating instances and provenance. It enjoys a high degree of interpretability: its decisions are supplied with a small relevant part of $D$,  which can be regarded as their explanation.

We implement and evaluate this candidate selection algorithm on a subset of BIRD-DEV
against three natural baselines. The experiments show that our method significantly outperforms
the baselines in the setting when a reliable consistency score is not available. We speculate that this is due to additional knowledge imparted to a general selection agent in form of $D'$ and that many existing selection agents can benefit from this additional information.

\begin{example}
Suppose that we need to select between 
\begin{quote}
SELECT COUNT(*) FROM account INNER JOIN district ON account.district\_id = district.district\_id WHERE account.frequency = 'POPLATEK PO OBRATU' AND district.a3 = 'East Bohemia' 
\hspace{1cm} $(Q_1)$    
\end{quote}
and
\begin{quote}
SELECT COUNT(*) FROM account INNER JOIN district ON account.district\_id = district.district\_id 
WHERE account.frequency = 'POPLATEK PO OBRATU' AND district.a3 = 'east Bohemia'
\hspace{1cm} $(Q_2)$    
\end{quote}
(which are different only in the capitalization of the penultimate word)
in context of Q\_nl = ``How many accounts who choose issuance after transaction are staying in East Bohemia region?''. First, both queries are executed on the model database $D$. As the correct spelling of this region in $D$ is ‘east Bohemia’, $Q_1(D) = 0$  while $Q_2(D) > 0$. By picking a pair of rows that 
contributes to (the before-counting part of) $Q_2(D)$
we obtain following separating instance $D’$:
\begin{verbatim}
{ 
  'row_0_of_account': {
    'account_id': 3837, 'district_id': 48, 
    'frequency': 'POPLATEK PO OBRATU’
  },
  'row_0_of_district': {
    'district_id': 48, 
    'a3': 'east Bohemia’
  }
}
\end{verbatim}
that has two tables, account and district, each of them has a single row, and these two rows share the same district\_id. When Q\_nl is evaluated on this instance, the answer is 1, 
meanwhile $Q_1(D’) = 0$, $Q_2(D’) = 1$, and so we conclude that $Q_2$ is correct and $Q_1$ is not. Note that this decision was made after we looked for both values 'East Bohemia' and 'east Bohemia' in database and discovered that only one of them occurs there.
\end{example}

\begin{example}
Suppose that we need to select between
\begin{quote}
SELECT loan.loan\_id, district.a3 AS district, district.a11 AS average\_salary FROM loan INNER JOIN account ON loan.account\_id = account.account\_id INNER JOIN district ON account.district\_id = district.district\_id WHERE loan.duration = 60 \hspace{1cm} $(Q_1)$    
\end{quote}
and
\begin{quote}
SELECT t1.loan\_id, t3.a2, t3.a11 FROM loan AS t1 INNER JOIN account AS t2 ON t1.account\_id = t2.account\_id INNER JOIN district AS t3 ON t2.district\_id = t3.district\_id WHERE t1.duration = 60
\hspace{1cm} $(Q_2)$    
\end{quote}
given the question "List the loan ID, district and average salary for loan with duration of 60 months", the slightly misleading hint “A3 refers to regions; A11 refers to average salary” and the (relevant part of the) schema
\begin{verbatim}
CREATE TABLE `district` (
	`district_id` -- location of branch
	`A2` TEXT, -- district_name 
	`A3` TEXT, -- region 
	`A11` INTEGER, -- average salary
);
\end{verbatim}
As these queries are different only in the second term of the 'select' list,
by tracking the computation of the common conjunctive part of $Q1$ and $Q2$ we 
obtain the following  separating database instance $D’$ with single-row tables:
\\[2mm]
\noindent
\hbox to \columnwidth{%
  \vbox{%
    \hbox{\hskip 4mm\textbf{account}\strut}
    \vskip 1mm%
    \hbox{%
      \begin{tabular}{@{}cc@{}}
        \hline account\_id & district\_id \\ \hline 5181 & 75 \\ \hline
      \end{tabular}%
    }%
  }%
  \hfil
  \vbox{%
    \hbox{\hskip 4mm\textbf{loan}\strut}%
    \vskip 1mm%
    \hbox{%
      \begin{tabular}{@{}ccc@{}}
        \hline loan\_id & account\_id & duration \\ \hline 6055 & 5181 & 60 \\ \hline
      \end{tabular}%
    }%
  }%
}
\vskip 1.5ex\relax%
%
\noindent
\vbox{%
  \textbf{district}\strut \\
  \begin{tabular}{cccc}
    \hline district\_id & a2 & a3 & a11 \\ \hline 75 & Prerov & north Moravia & 8819 \\ \hline
  \end{tabular}%
}\\[2mm]
When Q\_nl is executed on $D’$, the answer is [[6055, "Prerov", 8819]] (here the LLM could have used its domain knowledge), meanwhile $Q_1$ gives [[6055, "north Moravia", 8819]], $Q_2$ gives [[6055, "Prerov", 8819]], and so we conclude that 
$Q_2$ is correct and $Q_1$ is not.
\end{example}

\begin{example}
Suppose that we need to select between 
\begin{quote}
SELECT AVG(loan.amount) FROM client INNER JOIN disp ON client.client\_id = disp.client\_id INNER JOIN loan ON disp.account\_id = loan.account\_id WHERE client.gender = 'M' \hspace{1cm} $(Q_1)$
\end{quote}
and
\begin{quote}
SELECT AVG(loan.amount) FROM loan INNER JOIN disp ON loan.loan\_id = disp.disp\_id INNER JOIN client ON disp.client\_id = client.client\_id WHERE client.gender = 'M' \hspace{1cm} $(Q_2)$    
\end{quote}
given the question “What is the average loan amount by male borrowers”.
The separating instance D’ will be\\[2mm]
\noindent
\hbox to \columnwidth{%
  \vbox{%
    \hbox{\textbf{client}\strut}%
    \vskip 1mm%
    \hbox{%
      \begin{tabular}{cc}
        \hline
        client\_id & gender \\
        \hline
        5117 & M \\
        9505 & M \\
        \hline
      \end{tabular}%
    }%
  }%
  \hfil%
  \vbox{%
    \hbox{\textbf{disp}\strut}%
    \vskip 1mm%
    \hbox{%
      \begin{tabular}{ccc}
        \hline
        disp\_id & client\_id & account\_id \\
        \hline
        5117 & 5117 & 4245 \\
        9197 & 9505 & 7674 \\
        \hline
      \end{tabular}%
    }%
  }%
}%
\vskip 1.5ex\relax
%
\noindent
\vbox{%
  \textbf{loan}\strut \\
  \begin{tabular}{ccc}
    \hline
    loan\_id & account\_id & amount \\
    \hline
    5117 & 718 & 76944 \\
    6562 & 7674 & 94488 \\
    \hline
  \end{tabular}%
}\\[2mm]

It has three tables with two rows in each table, and the rows correspond to the correct and the incorrect ways of joining the rows. When Q\_nl is executed on $D’$, the LLM makes a correct join and ignores incorrect, and produces the answer [[94488]]. When $Q_1$ and $Q_2$ are evaluated on $D’$, they give respectively [[94488]] and [[76944]], so we conclude that $Q_1$ 
is correct while $Q_2$ is not.
\end{example}

\section{Candidate Selection Algorithm}

\begin{figure}
\centerline{\scalebox{0.78}{\includegraphics{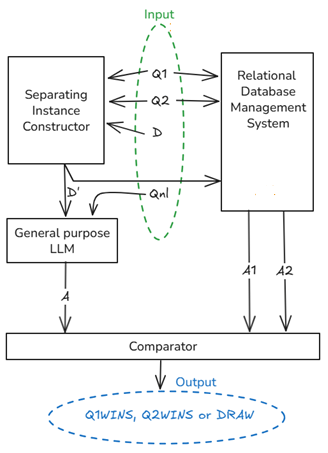}}}
\caption{Binary selection unit} \label{fig1}
\end{figure}

\paragraph{Tournament Selection}
We select an optimal SQL query in a two-stage process. First, we consolidate an initial pool of queries by grouping them based on identical execution results on a database $D$. 
A single representative is chosen from each group to form a candidate set
$C' = \{c_1, \dots , c_m\}$. Second, these candidates compete in a pairwise round-robin tournament using a binary selection unit shown in Figure~\ref{fig1}.  

The crucial part is the separating instance constructor which receives the candidates 
$Q_1$ and $Q_2$ and the model database $D$ and produces $D’$ 
which is a subset of $D$ such that that $Q_1$ and $Q_2$ produce on $D’$ different answers.  
Then $D’$ and Q\_nl are fed into a general purpose LLM together with the database schema, 
the hint and metadata, and the LLM is asked what it thinks is the answer to Q\_nl on the database instance $D’$.  At the same time $Q_1$ and $Q_2$ are evaluated on $D’$ using a standard relational database management system thus producing the answers $A_1$ and $A_2$. Then the decision on binary selection is made in the comparator, which returns Q1WINS if $A = A1  \neq A2$, 
Q2WINS if $A = A_2 \neq A_1$ and DRAW otherwise.

As usual, a win gives 1 point, a loss 0 points and a draw 0.5 points. Finally, the winner of the tournament (a candidate with the greatest score) is selected. If there are multiple winners, 
we fall back on the consistency score.

\paragraph{Separating Instance Constructor}
\emph{Provenance tokens} are unique labels associated to the rows of the tables in $D$. 
A simple term is a $\otimes$-product of provenance tokens. 
A general provenance term in addition to $\otimes$ can also contain $\oplus$ and $\ominus$ operations (see \cite{sen2025provsql} for a precise definition).
Now we describe a naive heuristic algorithm for constructing separating instances,
on which we settled in our experiments:

\begin{description}
\item{\bf Input:} $Q_1$, $Q_2$, $D$ 
\item{\bf Assumption:} $Q_1(D)  \neq Q_2(D)$
\item{\bf Returns:} a subset $D'$ of $D$ such that\\ 
\hspace*{3cm}$Q_1(D') \neq Q_2(D')$ or {\bf failure}
\end{description}

\begin{itemize}
\item Evaluate $Q_1$ and $Q_2$ on $D$ using RDBMS with provenance support.
\item Let $A_1$ and $A_2$ be the answers produced by $Q_1$ and $Q_2$ on $D$.
\item Parse the provenance terms for $A_1$ and $A_2$ into the sets $P_1$ and $P_2$
of simple terms (or return {\bf failure}) 
\item If $P_1 = P_2$, repeat 42 times:
\begin{itemize}
\item pick two random terms from $P_2 = P_1$ and set $D'$ to be generated by these simple terms.
\item stop and return $D'$ if $Q_1$ and $Q_2$ give different answers on $D'$
\end{itemize}
\item If $P_1 \neq P_2$, repeat 42 times:
\begin{itemize}
\item set $ M, N = 
\begin{cases} 
(3,4) & \text{if } Q_1 \text{ or } Q_2 \text{ contain words} \\ 
      & \text{\hspace*{5mm}`sum', `case' and `when'}\\
(1,2) & \text{elif } Q_1 \text{ or } Q_2 \text{ contain `count'} 
\\ (1,1) & \text{otherwise} 
\end{cases} $
\item pick $M$ random terms $P_1 - P_2$ and $N$ random terms from $P_2 - P_1$  
and set $D'$ to be generated by these simple terms.
\item stop and return $D'$ if $Q_1$ and $Q_2$ give different answers on $D'$
\end{itemize}
\item Return {\bf failure}
\end{itemize}    

A failure results in a draw in the tournament table.

\paragraph{Evaluation of queries in natural language on small instances}
\label{nl-ev}
We use Qwen3-Coder-30B-A3B-Instruct with default parameters 
and the prompt from Appendix~\ref{prompt-base}
to evaluate $Q_{nl}$ on the small separating instance. As a rule the answers are
correct, but not always. To illustrate, suppose we want to know
`Who are the female account holders who own credit cards and also have loans?' given
`Female refers to gender = 'F'\ ' on the following small separating instance:
{\footnotesize
\begin{verbatim}
### client                ### card               

| client_id | gender |    | disp_id | type     |    
|-----------|--------|    |---------|----------|    
| 2063      | F      |    | 2063    | classic  |   
| 9675      | F      |                              

### loan

| account_id |
|------------|
| 1698       |
| 7824       |

### disp

| disp_id | client_id | account_id | type   |
|---------|-----------|------------|--------|
| 2063    | 2063      | 1698       | OWNER  |
| 9367    | 9675      | 7824       | OWNER  |
\end{verbatim}
}
Appendix~\ref{output-base-correct} shows a typical output of the LLM with 
the correct answer [[2063]]. However, sometimes 
(in 35\% cases for instances in json and 23\% cases for instances in markdown)
the LLM produces incorrect answer [[2063], [9675]], since it mistakenly
comes to conclusion that both clients own cards.

These hallucinations are the bottleneck of our selection method. To mitigate
their effect, 
we carry out evaluation of $Q_{nl}$ three times and apply majority voting. 
This modification is also applied to the baselines to have a fair comparison.

\section{Experiments}

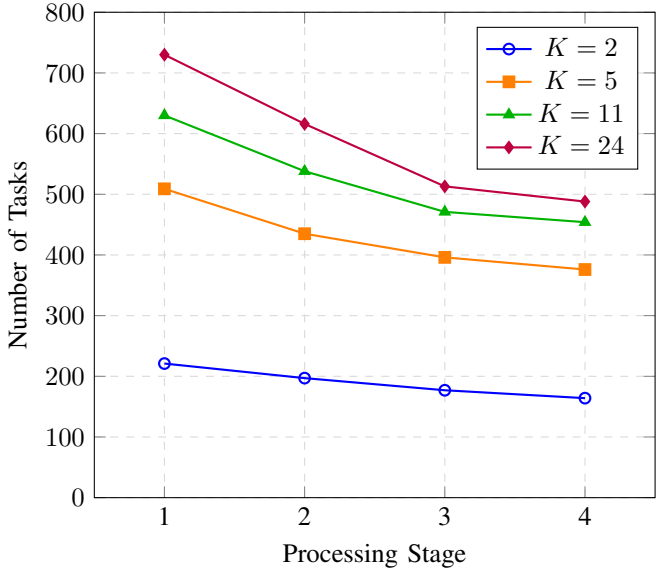
\begin{figure}
\begin{tikzpicture}
\begin{axis}[
    width=9cm,
    height=8cm,
    xlabel={Processing Stage},
    ylabel={Number of Tasks},
    xmin=0.5, xmax=4.5,
    xtick={1,2,3,4},
    xticklabels={1,2,3,4},
    ymin=0, ymax=800,
    ytick distance=100,
    legend pos=north east,
    grid=both,
    grid style={dashed, gray!30},
    ymajorgrids=true,
    xmajorgrids=true,
    nodes near coords align={above},
    every node near coord/.append style={font=\small, /pgf/number format/fixed, /pgf/number format/precision=0},
]

\addplot[blue, mark=o, thick] coordinates {
    (1, 221)
    (2, 197)
    (3, 177)
    (4, 164)
};
\addlegendentry{$K=2$}

\addplot[orange, mark=square*, thick] coordinates {
    (1, 509)
    (2, 435)
    (3, 396)
    (4, 376)
};
\addlegendentry{$K=5$}

\addplot[green!70!black, mark=triangle*, thick] coordinates {
    (1, 630)
    (2, 538)
    (3, 471)
    (4, 454)
};
\addlegendentry{$K=11$}

\addplot[purple, mark=diamond*, thick] coordinates {
    (1, 730)
    (2, 616)
    (3, 513)
    (4, 488)
};
\addlegendentry{$K=24$}

\end{axis}
\end{tikzpicture}
\caption{Task filtering pipeline.}\label{fig2a}
\end{figure}

\paragraph{Dataset Preparation}
We started with the original version of the BIRD-DEV dataset~\cite{bird_sql} 
comprising 1,534 NL2SQL tasks in which we manually corrected around 50 obviously incorrect ground answers. For each task, we generated SQL candidates using AlphaSQL~\cite{li2025alpha} with rollout counts $K \in \{2, 5, 11, 24\}$
and clusterized them according to their execution results on model databases. 
At stage 1 the tasks where all candidates give the same answer and the tasks without correct candidates were removed. From the rest we kept only the tasks where all candidate representatives could be translated into PostgreSQL dialect (stage 2), executed on the ProvSQL~\cite{sen2025provsql} extension of Postgres RDBMS (stage 3), and of which the provenance terms 
were successfully parsed (stage 4). This gave us four datasets comprising 164, 376, 454 and 488 tasks. The number of tasks on each of the stages is shown in Figure~\ref{fig2a}.

\begin{figure}
\begin{tikzpicture}
\begin{axis}[
    width=9cm,
    height=8cm,
    xlabel={Number of Rollouts},
    ylabel={Accuracy (\%)},
    xmode=log,
    xmin=1.5, xmax=30,
    xtick={2,5,11,24},
    xticklabels={2,5,11,24},
    ymin=50, ymax=80,
    ytick={50,55,60,65,70,75,80},
    legend pos=south east,
    grid=both,
    grid style={dashed, gray!30},
    ymajorgrids=true,
    xmajorgrids=true,
    nodes near coords align={above},
    every node near coord/.append style={font=\small, /pgf/number format/fixed, /pgf/number format/precision=1},
    xticklabel style={/pgf/number format/fixed},
]

\addplot[blue, mark=o, thick] coordinates {
    (2, 54.3)
    (5, 72.6)
    (11, 75.1)
    (24, 71.7)
};
\addlegendentry{Consistency}

\addplot[orange, mark=square*, thick, nodes near coords align={below}] coordinates {
    (2, 63.9)
    (5, 71.3)
    (11, 71.1)
    (24, 74.5)
};
\addlegendentry{Naive}

\addplot[green!70!black, mark=triangle*, thick] coordinates {
    (2, 64.3)
    (5, 77.8)
    (11, 77.5)
    (24, 79.0)
};
\addlegendentry{DeepEye}

\addplot[red, mark=diamond*, thick, nodes near coords align={below}] coordinates {
    (2, 77.9)
    (5, 76.9)
    (11, 75.3)
    (24, 78.5)
};
\addlegendentry{Ours}

\end{axis}
\end{tikzpicture}
\caption{Evaluation results.}
\label{fig2b}
\end{figure}
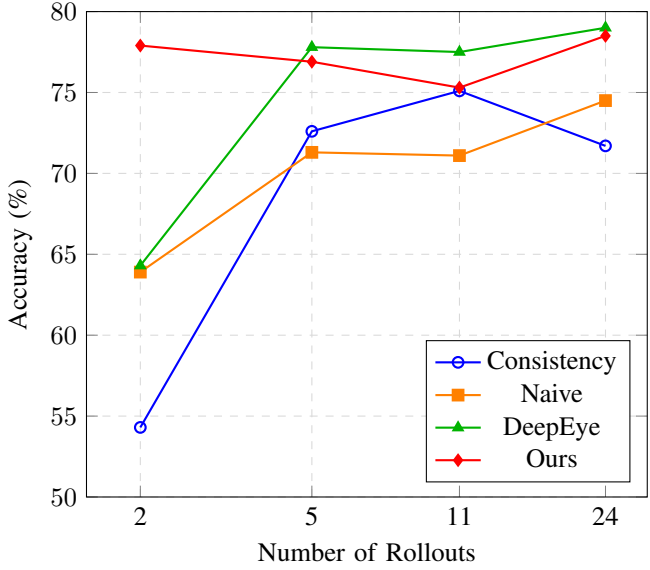

\paragraph{Baselines}
We compare our approach against three baselines: the \emph{Consistency} score from AlphaSQL~\cite{li2025alpha}, a \emph{Naive} selection strategy from an unofficial AlphaSQL 
implementation (see Appendix~\ref{naive}), and the consistency-aware candidate selection module from DeepEye-SQL~\cite{li2025deepeye}.

\paragraph{Evaluation}
For each of the four approaches to selection we measured execution accuracy 
on our four datasets and plotted it in Figure~\ref{fig2b}. 
The results for baselines are averaged over 4 runs and the results of our method over 16 runs to take into account the randomness of LLMs. The sets of candidates (and so the consistency baseline) were calculated for each $K$ only once. In all our calculation
Qwen3-Coder-30B-A3B-Instruct with default parameters  \{"repetition\_penalty": 1.05, "temperature": 0.7, "top\_p": 0.8, "top\_k": 20\} was used as the driving LLM.

\paragraph{Coverage}
The \emph{technical coverage} on BIRD-DEV of our method can be estimated as the ratio 
between the number of tasks on stage 4 to the number of tasks on stage 1. In our experiments
it declines from 74\% for $K = 2$ to 67\% for $K=24$. The \emph{actual coverage} takes
out from the technical coverage the tasks with $\ge 2$ tournament winners 
and is around 50\% for $K=24$. The \emph{success rate} of our separating instance constructor on all candidate pairs for this value of $K$ is 95\%.

\paragraph{Comparing answers}
We remove from the answers entirely null rows, then remove duplicates, 
and then compare thus simplified answers in a row-by-row manner by comparing its
rows as sets. This is due to the fact that in most BIRD tasks it is unclear if
the user wants the answer with 'DISTINCT' or 'NOT NULL' modifiers, and what 
order of columns they prefer.

\paragraph{Dealing with limitations of ProvSQL}
ProvSQL extension of Postgres has many limitations, the most notorious
of which is that it does not support subqueries in WHERE-clauses. To deal 
with it, firstly, when we select representatives, we try to pick
one which does not have subqueries in WHERE-clauses. In one of our 
early experiments this increased
the number of valid tasks from 314 (out of 1534) to 530. If this is not possible,
we automatically reformulate the queries by transforming subqueries in WHERE-clauses
into joins. This increased the number of valid tasks further to 594.

\paragraph{Representing instances}
We considered three ways of representing small database instances when passing them to the LLM:
SQL (insert statements), markdown (as in Section~II-c) 
and json (as in Example 1).  In a specially designed experiment to compare the representation
over a range of tasks json showed the best result, 
then goes markdown, then SQL. So we stuck to json in our final evaluation.

\section{Explanations of decisions and\\evidence-based criticism of gold answers}

An additional advantage of the suggested selection technique 
is that the selection result is explained by 
tournament reports which often can be readily transformed into an evidence-based
criticism of gold answers. `Evidence' here consists of a few concrete
rows of the model database together with an informal argument
in favour of the alternative answer. Here is an example:\\[2mm]
BIRD ENTRY
{\footnotesize
\begin{verbatim}
{
    "question_id": 376,
    "db_id": "card_games",
    "question": "What is the card layout 
             of cards with keyword of flying?",
    "evidence": "",
    "SQL": "SELECT layout FROM cards 
                 WHERE keywords = 'Flying'",
    "difficulty": "simple"
}
\end{verbatim}
}
CRITIQUE
{\footnotesize
\begin{verbatim}
Table 'cards' contains a row with 
{
      "id": 25069,
      "name": "Reidane, God of the Worthy",
      "keywords": "Flying,Vigilance",
      "layout": "modal_dfc"
}

This layout should be included in the answer, but 
the 'gold' answer does not  return it. The query

SELECT DISTINCT cards.layout AS layout FROM cards 
AS cards WHERE cards.keywords LIKE '%Flying%'

seems better than the 'gold' answer.
\end{verbatim}
}

There are more examples of such reports in Appendix~\ref{evidence}. 

\section{Conclusion}

The starting point of this paper is the observation that the majority of
existing candidate selection strategies have
limited access to the values contained in the database. 
This can be to an extent mitigated by database preprocessing 
(e.g., \cite{alpha-sql-preprocessing}),
which allows to insert task-specific examples of values into schema description.
Here we made an attempt to increase the data-awareness of the selection
unit by supplying it with a separating instance which is extracted from the model database and 
so inherits all its values and universal constraints  (cf.~\cite{klopfenstein2026spotit+}). 
Think of a separating instance as a `magnifying glass' which retrieves from a large database a small portion which is crucial for making an
informed decision between the two candidates. 

A broader contribution of this paper is an original framework for combining symbolical reasoning with LLM-based inference. Our heuristic algorithm for constructing  separating instances achieves 95\% success rate in our setting
but lacks theoretical guarantees, currently serving as a placeholder for more rigorous symbolic procedures. Thus, we highlight the extraction of a minimal separating instance from a database to distinguish between two given SQL queries as a logical problem of utmost practical importance (cf. \cite{miao2019explaining}).

Our third contribution is a collection of SQL candidates for original 
BIRD questions for benchmarking candidate selection available at
\url{http://github.com/staskikotx/SISelection/tree/main/paper/candidates}.

The adoption of the flowchart from Figure~\ref{fig1} for experiments
is somewhat arbitrary --- 
in hindsight, a simpler architecture could have given a better result --- but 
we went with it due to aesthetic considerations. This solution was implemented
and evaluated against three natural baselines using candidates produced by AlphaSQL
with number of rollouts $K \in \{2,5,11,24\}.$ We discovered that in case $K = 2$ (the setting we observed in a Text2Report chatbot kindly shown to us by Erwin Hurtado) it 
significantly surpasses the baselines. In other cases it surpasses `Naive' selection, on a par with the consistency baseline and is second to DeepEye consistency-aware selection unit, probably, because our method is more prone to hallucinations. Dealing with this effect 
by prompt-engineering, changing the architecture of the selection unit, changing/fine-tuning the underlying LLM or some other means we leave as a topic for further research.

%
%
\bibliographystyle{IEEEtran}
\bibliography{sql}
\appendices
\section{Prompt Template for Base Algorithm.}\label{prompt-base}
\begin{lstlisting}[style=promptstyle]
You are an experienced database expert. You need to evaluate a query in natural language into a small set of tuples of values, given the database  information, the database instance, a question and some additional  information. The database structure is defined by the following table schemas (comments after '--' provide additional column descriptions). Note that the "Value Examples" are actual values from the column.  Some column might contain the values that are directly related  to the question. Use it to help you justify which columns to use.

Given the table schema description and the `Question`, you need to understand  the database and columns. First you need to understand what information the  user who submitted the question wanted to get. Then read the code of  the database instance, think, and output the tuples which constitute  the answer to the `Question`. All your tuples should be of length  {tuple_length} . Please give you row-by-row workout where for  each table row you print the values of attributes you select   and your calculations and decisions regarding the filters in WHERE-clauses.

**************************
[Database schema]
{schema}
**************************
[Question] 
{question}

[Hint]
{hint}

[Database Instance]:
{instance}
**************************
Reread the question, the hint, the database schema and the database instance. Think step by step and output the list of required tuples of values. The most important input data is the 'Database Instance' and the 'Question'. Take a deep breath and think logically. Do not be too helpful, be smart. If you are given two items, it is likely that you have to select one. If you are given one item, it is likely that you need to make a decision whether select it or not. If not, you need to return an empty answer [] as set of tuples. If you are asked to select entities from a certain table (such as card, people or withdrawals), please return  their ids rather than names or other values. When you join rows from  different tables, pay attention at the names of the attributes and their  explanations in schema.

Please respond with a valid JSON object structured as follows:

```json
{{
  "chain_of_thought_reasoning": 
          "Your reasoning for evaluating the SQL query",
  "workout": "row-by-row calculations for each of insert statements",
  "question": "question in natural language",
  "tuples_that_answer_question": 
           [[value1, value2, ... ], [value3, value4, ...] ... ]
}}
```

Use " for string literals and reals. Sometimes you will have integer  numbers instead of values, do not use " or ' for them. If your answer is  a single integer number, keep it in double brackets, like this:   "tuples": [[200]]

Only output a json as your response. 
\end{lstlisting}

\section{Typical LLM Output.}\label{output-base-correct}
\begin{lstlisting}[style=promptstyle]
```json
{
  "chain_of_thought_reasoning": "To answer the question, I need to identify female account holders who own credit cards and also have loans. Let me break this down step by step:  1. First, I need to find female clients (gender = 'F') 2. Then, I need to find those who own credit cards (through the card table) 3. And also have loans (through the loan table)  Looking at the database instance: - Client 2063 is female and has disp_id 2063 which maps to account 1698 - Client 9675 is female and has disp_id 9367 which maps to account 7824 - Both accounts (1698 and 7824) have loans - Both clients have credit cards (disp_id 2063 shows type 'classic', disp_id 9367 is not shown but we need to check if it has a card)  Wait, let me reconsider - the card table only shows disp_id 2063 with type 'classic'. This means only client 2063 has a card. But client 9675's disp_id 9367 is not in the card table, so client 9675 doesn't have a card.  So I need to find female clients who: 1. Have a card (disp_id in card table) 2. Have an account with a loan  From the instance: - Client 2063 is female, has a card (disp_id 2063), and has loan on account 1698 - Client 9675 is female but does not have a card (disp_id 9367 not in card table)  Therefore, only client 2063 meets all criteria.",
  "workout": "Step 1: Identify female clients from client table where gender = 'F' - client_id 2063, gender 'F' - client_id 9675, gender 'F'  Step 2: Find which of these female clients own credit cards - From card table: disp_id 2063 has type 'classic' - From disp table: client 2063 has disp_id 2063 - So client 2063 owns a credit card - Client 9675 has disp_id 9367 which is NOT in card table, so does not own a credit card  Step 3: Check if these clients have loans - For client 2063: account_id 1698 has loan (in loan table) - For client 9675: account_id 7824 has loan (in loan table)  Step 4: Combine conditions - Client 2063: female, owns credit card, has loan -> meets criteria - Client 9675: female, does NOT own credit card, has loan -> does not meet criteria  Final selection: Only client_id 2063 meets all conditions",
  "question": "Who are the female account holders who own credit cards and also have loans?",
  "tuples_that_answer_question": [[2063]]
}
```
\end{lstlisting}

\section{A prompt for 'naive' selection}\label{naive}

\begin{lstlisting}[style=promptstyle]
Given a DB schema, a question, and a knowledge evidence, and some SQLite queries expressing the question are given, please choose the most accurate SQL based on the Checklist.

### Instructions:
- If the first SQL satisfies all the conditions of the checklist, please choose the first SQL. If not, move on to the next SQL.
- If there is no SQL that satisfies all the requirements on the checklist, just choose the first SQL.
- First, think about the principles involved in solving this question which could be helpful.
- Then, provide a detailed step-by-step explanation following the order of the checklist when checking whether each SQL satisfies the checklist.

### Checklist:
1. The SQL should accurately represent the question.
2. The SQL should accurately use the given knowledge evidence.
3. The SELECT clause should not include any additional columns that are not included in the question.
4. The order of column(s) in the SELECT clause must be the same as the order in the question.
5. Check if the operations are being performed correctly according to the column type.

### Database Schema:
{DB_SCHEMA}

### Question:
{QUESTION}

### Knowledge Evidence:
{KNOWLEDGE_EVIDENCE}

### Candidate SQL Queries:
{SQL_QUERIES}

### Your answer should strictly follow the following json format:
```json
{{
  "principles": "",  // The principles involved in solving this question.
  "reasoning": "",  // The reasoning steps for choosing the best SQL.
  "sql": "",  // The final chosen SQL.
}}
```

### Your Answer:
\end{lstlisting}

\section{Examples of reports\\ on likely incorrect 'gold' answers}\label{evidence}

BIRD-DEV-425

{\footnotesize
\begin{verbatim}
{
  "question_id": 425,
  "db_id": "card_games",
  "question": "What are the card numbers that don't 
      have multiple faces on a single card and have 
      the subtypes Angel and Wizard?",
  "evidence": "don't have multiple 
      faces on a single card side is null",
  "SQL": "SELECT id FROM cards WHERE 
      subtypes = 'Angel,Wizard' AND side IS NULL",
  "difficulty": "simple"
}
\end{verbatim}
}

CRITIQUE

'Card number' likely refers not to 'id', but to 'number' column of the 
'card' table. This column contains reasonable values, 
so probably they should be returned, not the id:

{\footnotesize
\begin{verbatim}
  id   |           name           | number
-------+--------------------------+-------
 17233 | Eater of Days            | 120
 25380 | Tergrid, God of Fright   | 307
 17480 | Segmented Krotiq         | 202
\end{verbatim}
}

Note that if `number` column contained NULLs, selecting ids would be OK.

\bigskip
\hrule
\bigskip

BIRD-DEV-581

{\footnotesize
\begin{verbatim}
{
  "question_id": 581,
  "db_id": "codebase_community",
  "question": "Who is the editor of the post titled 
      'Open source tools for visualizing 
      multi-dimensional data?'",
  "evidence": "'Open source tools for visualizing 
      multi-dimensional data' is the Title of Post; 
      editor refers to DisplayName;",
  "SQL": "SELECT T2.DisplayName FROM posts AS T1 
      INNER JOIN users AS T2 ON T1.OwnerUserId = T2.Id 
      WHERE T1.Title = 'Open source tools for 
          visualizing multi-dimensional data?'",
  "difficulty": "moderate"
},
\end{verbatim}
}

CRITIQUE

{\footnotesize
\begin{verbatim}
The database contains the following rows:
"row_0_of_posts_table": {
  "id": 196,
  "posttypeid": 1,
  "acceptedanswerid": 232,
  "owneruserid": 87,
  "lasactivitydate": "2012-11-21T06:38:02+08:00",
  "title": "Open source tools for visualizing 
      multi-dimensional data?",
  "lasteditoruserid": 9007,
  "lasteditdate": "2012-11-21T06:25:07+08:00",
  "communityowneddate": "2010-07-20T02:35:32+08:00",
},
"row_0_of_users_table": {
  "id": 9007,
  "reputation": 1318,
  "creationdate": "2012-02-07T01:02:53+08:00",
  "displayname": "naught101"
},
"row_1_of_users_table": {
  "id": 87,
  "reputation": 740,
  "creationdate": "2010-07-19T19:34:45+08:00",
  "displayname": "Paul"
}
  
Row_0_of_posts_table and row_0_of_users_table are 
connected by "lasteditoruserid" = 9007. So the editor 
of this post should be 'naught101', rather than 'Paul' 
who is its owner. 
\end{verbatim}
}

\bigskip
\hrule
\bigskip

BIRD-DEV-755

{\footnotesize
\begin{verbatim}
{
  "question_id": 755,
  "db_id": "superhero",
  "question": "List down at least five full name of 
      Demi-God superheroes.",
  "evidence": "Demi-God superheroes refers to 
      race = 'Demi-God'",
  "SQL": "SELECT T1.full_name FROM superhero AS T1 
      INNER JOIN race AS T2 ON T1.race_id = T2.id 
      WHERE T2.race = 'Demi-God'",
  "difficulty": "simple"
},
\end{verbatim}
}

CRITIQUE

{\footnotesize
\begin{verbatim}

The database contains the following fragment:

 "instance_as_dict": {
    "row_0_of_race_table": {
      "id": 16,
      "race": "Demi-God"
    },
    "row_0_of_superhero_table": {
      "id": 407,
      "superhero_name": "Kratos",
      "full_name": "-",
      "gender_id": 1,
      "race_id": 16,
      "publisher_id": 25,
      "height_cm": 198,
      "weight_kg": 108
    }
  }

which produces '-' in the 'gold' answer. The answer 
produced by a query with extra clause 
'WHERE full_name <> '-' seems better. 
\end{verbatim}
}
Again, this critique makes sense only in context of the model database.

\end{document}